\documentclass{aastex61}

\newcommand\aastex{AAS\TeX}

\submitjournal{ApJ}

\shorttitle{\aastex\ FSR~1716 Orbit}
\shortauthors{R. Contreras Ramos et al.}

\begin{document}

\title{THE ORBIT OF THE NEW MILKY WAY GLOBULAR CLUSTER FSR1716 = VVV-GC05\footnote{Based  on  observations  taken  within  the  ESO  programmes 179.B-2002 and  298.D-5048}}

\correspondingauthor{Rodrigo Contreras Ramos}

\author{Rodrigo Contreras Ramos}
\affiliation{Instituto Milenio de Astrof\'isica, Santiago, Chile.}
\affiliation{Pontificia Universidad Cat\'olica de Chile, Instituto de Astrof\'isica, Av. Vicu\~na Mackenna 4860, 7820436 Macul,  Santiago, Chile.}

\author{Dante Minniti}
\affiliation{Departamento de Ciencias F\'isicas, Facultad de Ciencias Exactas, Universidad Andr\'es Bello, Av. Fern\'andez Concha 700, Las Condes, Santiago, Chile.}
\affiliation{Vatican Observatory, V00120 Vatican City State, Italy.}
\affiliation{Instituto Milenio de Astrof\'isica, Santiago, Chile.}

\author{Jos\'e G. Fern\'andez-Trincado}
\affiliation{Departamento de Astronom\'\i a, Casilla 160-C, Universidad de Concepci\'on, Concepci\'on, Chile}	 
\affiliation{Institut Utinam, CNRS UMR 6213, Universit\'e Bourgogne-Franche-Comt\'e, OSU THETA Franche-Comt\'e, Observatoire de Besan\c{c}on, BP 1615, 25010 Besan\c{c}on Cedex, France}  

\author{Javier Alonso-Garc\'ia}
\affiliation{Unidad de Astronom\'ia, Facultad Cs. B\'asicas, Universidad de Antofagasta, Avda. U. de Antofagasta 02800, Antofagasta, Chile.}
\affiliation{Instituto Milenio de Astrof\'isica, Santiago, Chile.}

\author{M\'arcio Catelan}
\altaffiliation{On sabbatical leave at European Southern Observatory, Av. Alonso de C\'ordova 3107, 7630355 Vitacura, Santiago, Chile.}
\affiliation{Instituto Milenio de Astrof\'isica, Santiago, Chile.}
\affiliation{Pontificia Universidad Cat\'olica de Chile, Instituto de Astrof\'isica, Av. Vicu\~na Mackenna 4860, 7820436 Macul,  Santiago, Chile.}

\author{Felipe Gran}
\affiliation{Instituto Milenio de Astrof\'isica, Santiago, Chile.}
\affiliation{Pontificia Universidad Cat\'olica de Chile, Instituto de Astrof\'isica, Av. Vicu\~na Mackenna 4860, 7820436 Macul,  Santiago, Chile.}

\author{Gergely Hajdu}
\affiliation{Pontificia Universidad Cat\'olica de Chile, Instituto de Astrof\'isica, Av. Vicu\~na Mackenna 4860, 7820436 Macul,  Santiago, Chile.}
\affiliation{Astronomisches Rechen-Institut, Zentrum f\"ur Astronomie der Universit\"at Heidelberg, M\"onchhofstr. 12-14, 69120 Heidelberg, German}
\affiliation{Instituto Milenio de Astrof\'isica, Santiago, Chile.}

\author{Michael Hanke}
\affiliation{Astronomisches Rechen-Institut, Zentrum f\"ur Astronomie der Universit\"at Heidelberg, M\"onchhofstr. 12-14, 69120 Heidelberg, German}

\author{Maren Hempel}
\affiliation{Instituto Milenio de Astrof\'isica, Santiago, Chile.}
\affiliation{Pontificia Universidad Cat\'olica de Chile, Instituto de Astrof\'isica, Av. Vicu\~na Mackenna 4860, 7820436 Macul,  Santiago, Chile.}

\author{Edmundo Moreno D\'iaz}
\affiliation{Instituto de Astronom\'ia, Universidad Nacional Aut\'onoma de M\'exico, Apdo. Postal 70264, M\'exico D.F., 04510, M\'exico}

\author{\'Angeles P\'erez-Villegas}
\affiliation{Universidade de S\~ao Paulo, IAG, Rua do Mat\~ao 1226, Cidade Universit\'aria, 05508-900, S\~ao Paulo, Brazil}

\author{\'Alvaro Rojas-Arriagada}
\affiliation{Instituto Milenio de Astrof\'isica, Santiago, Chile.}
\affiliation{Pontificia Universidad Cat\'olica de Chile, Instituto de Astrof\'isica, Av. Vicu\~na Mackenna 4860, 7820436 Macul,  Santiago, Chile.}

\author{Manuela Zoccali}
\affiliation{Instituto Milenio de Astrof\'isica, Santiago, Chile.}
\affiliation{Pontificia Universidad Cat\'olica de Chile, Instituto de Astrof\'isica, Av. Vicu\~na Mackenna 4860, 7820436 Macul,  Santiago, Chile.}

\email{rcontrer@astro.puc.cl}


\begin{abstract}
We use deep multi-epoch near-IR images of the VISTA Variables in the V\'ia L\'actea (VVV) Survey to measure proper motions (PMs) of stars in the Milky Way globular cluster FSR1716 = VVV-GC05. The color-magnitude diagram of this object, made using PM selected members, shows an extended horizontal branch, nine confirmed RR Lyrae members in the instability strip, and possibly several hotter stars extending to the blue. Based on the fundamental-mode (ab-type) RR Lyrae stars that move coherently with the cluster, we confirmed that FSR1716 is an Oosterhoff I globular cluster with a mean period $\langle P_{ab} \rangle$=0.574 days. Intriguingly, we detect tidal extensions to both sides of this cluster in the spatial distribution of PM selected member stars. Also, one of the confirmed RRabs is located $\sim$ 11 arcmin in projection from the cluster center, suggesting that FSR1716 may be losing stars due to the gravitational interaction with the Galaxy. We also measure radial velocities (RVs) for five cluster red giants selected using the PMs. The combination of RVs and PMs allow us to compute for the first time the orbit of this globular cluster, using an updated Galactic potential. The orbit results to be confined within ${\rm |Z_{max}|<2.0}$ kpc, and has eccentricity $0.4<e<0.6$, with perigalactic distance ${\rm 1.5<R_{peri}(kpc)<2.3}$, and apogalactic distance ${\rm 5.3<R_{apo}(kpc)<6.4}$. We conclude that, in agreement with its relatively low metallicity ([Fe/H]$=-1.4$ dex), this is an inner halo globular cluster plunging into the  disk of the Galaxy. As such, this is a unique object to test the dynamical processes that contribute to the disruption of Galactic globular clusters.
\end{abstract}

\keywords{techniques: proper motions – Galaxy: kinematics and dynamics – Galaxy: disk – Galaxy: globular clusters: individual: FSR~1716 = VVV-GC05}

\section{Introduction} \label{sec:intro}
The Galactic globular clusters (GCs) can be destroyed by different dynamical processes, and we see today only a small fraction of the original population that managed to remain within the classical “survival triangle” \citep[e.g.][]{fall77,fall85,aguilar88}.
%
Isolated clusters are expected to evaporate along their lives due to internal stellar encounters  \citep{henon61,aarseth93}. 
%
In addition, the presence of the Galactic tidal field imposes a radial limiting size for a GC, that can vary along its orbit around the Galactic center. 
%
Dynamical friction is also important as the GC moves through an ocean of Galactic stars, acting preferentially in massive clusters moving in regions where the stellar density is very high, like the bulge or disk.
Interactions with the tidal field of the Galactic disk also cause shocks, that heat up the outer regions of GCs. The disk shocking occurs when the cluster crosses the Galactic disk, where it is compressed by the varying z-component of the Galactic potential \citep{leon00}, and this effect appears to dominate the heating of GCs \citep{chernoff86}. In addition, a GC may exhibit a transient deformation during the disk crossing \citep{leon00}. 

The star cluster FSR1716 = VVV-GC05 is a recently discovered Milky Way GC with a Heliocentric distance $D=7.5$ kpc that lies right in the Galactic plane, at $l =329.77812$, $b = -1.59227$ deg \citep{minniti17_1}, indicating that it may belong to the disk. However, the detection of RR Lyrae (RRL) members, coupled with the obtained metallicity, shows that this is an old and metal-poor GC, with age $>10$ Gyr, and metallicity ${\rm [Fe/H]}=-1.4$ dex \citep{minniti17_1,koch17}, an indication that it may belong to the Galactic halo. Depending on its orbit and considering its location and elongated shape \citep[$b/a=0.7$;][]{minniti17_1}, this cluster should be severely affected by dynamical processes, or even being presently disrupted. Therefore, this GC is a prime target for dynamical studies. 


In this paper we compute the orbit of VVV-GC05, based on our newly measured proper motions (PMs) (presented in section \ref{pm}) and radial velocities (presented in section \ref{rv}). We measure the orbital parameters of this cluster (section \ref{orbit}), which allow us to conclude that this is an inner halo Galactic GC (as discussed in section \ref{discussion}). In the future, it would be interesting to study to what extent the different dynamical processes are affecting VVV-GC05.

\section{THE VVV-GC05 PROPER MOTIONS} \label{pm}
The VISTA Variables in the V\'ia L\'actea Survey (VVV) is mapping the bulge and southern disk regions of the Milky Way in the near-IR. The $ZYJHK_{s}$ images and aperture photometry of this public survey  are available at the ESO Archive. The images are deep, reaching $K_s\sim17-18$ mag, and have high resolution (scale 0.34"/pix), allowing the measurement of accurate PMs \citep{libralato15,smith18,contrerasramos17}. In particular, \cite{contrerasramos17} measured the proper motions of the Milky Way GC NGC~6544 based on the VVV images collected between years 2010 and 2016, obtaining a clean sample of cluster stars down to two magnitudes below the main sequence turn off.

\begin{figure}[h]
\plotone{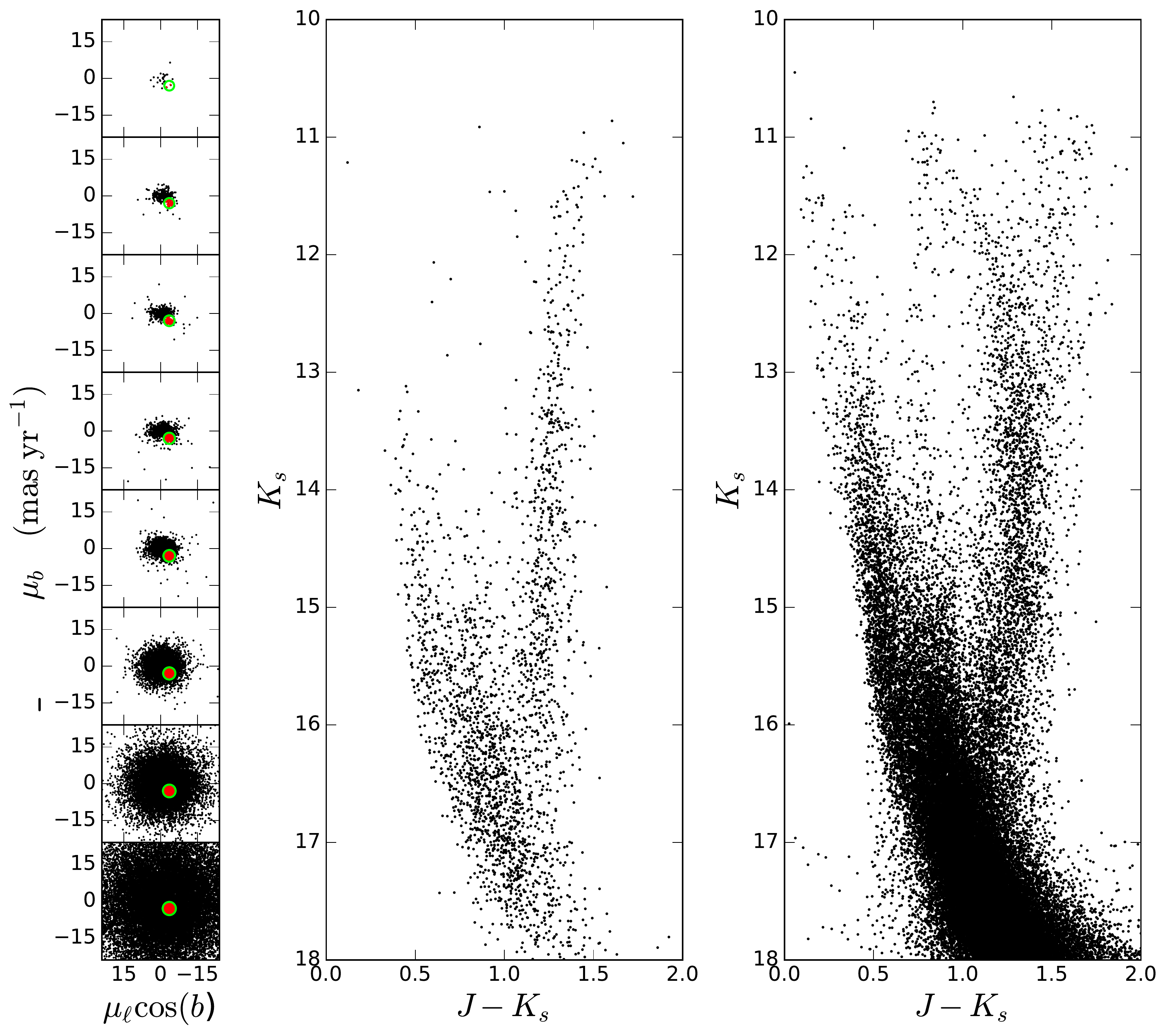}
\caption{VPD diagram (left) divided in eight one-magnitude bins, for all the stars detected in this work. The green circle located at ($\mu_{l}\cos(b), \mu_b = -3.48, -2.70$) mas~yr$^{-1}$ shows the region used to separate cluster (in red) from field stars, with a radius that varies from two mas (brightest stars) to three mas (faintest stars). CMD for stars that move coherently with the cluster (middle), and likely stars belonging to the disk (right) after cleaning cluster members. The PM selection of cluster members is still contaminated with some blue foreground disk stars, but the narrow RGB of the cluster is clearly defined.} 
\label{f_vpd_bin}
\end{figure}

We have measured the PMs of the stars in the field surrounding VVV-GC05, following the procedures described by \cite{contrerasramos17}. The main difference with respect to our previous work is that these PMs are measured relative to the blue plume made of disk stars in the cluster region (instead of bulge stars). The typical uncertainties are $\sim 1.18$ mas~yr$^{-1}$ in each PM coordinate, for stars with $K_s=15$ mag, 
accurate enough to compute orbital properties of the cluster. 
Figure \ref{f_vpd_bin} shows the relative PMs measured from the VVV images for a $12\times12$ arcmin field containing the GC VVV-GC05, where we have color coded the stars according to their density. This smooth vector point diagram (VPD) shows that a group of stars centered at $(\mu_{l}\cos(b),\mu_b)\sim(-4,-3)$ mas~yr$^{-1}$ clearly separates from the bulk of disk stars. The Galactic coordinates of this isolated group of stars correspond to the location of VVV-GC05 obtained by \cite{minniti17_1}. Taking the mean value for relatively bright stars ($K_s<15$) inside a 2 mas circle containing the cluster stars in the VPD, we measure the mean relative PM of GC VVV-GC05 in Galactic coordinates: $\mu_{l}\cos(b) = -3.74 \pm 0.1$ mas~yr$^{-1}$, and $\mu_b = -2.86 \pm 0.1$ mas~yr$^{-1}$. 
The same VPD but this time divided in eight one-magnitude bins is presented in Figure~\ref{f_vpd_bin} (left panel). The green circle in each plot shows the adopted membership criterion that we used to select cluster members (in red) in each magnitude interval. We adopted a radius of two mas~yr$^{-1}$ for the brightest stars with higher S/N ratios, on the top bin, relaxing the selection criteria for dimmer stars with lower S/N ratios (three mas~yr$^{-1}$ at the bottom). The resulting color-magnitude diagrams (CMDs) for both the stars selected as cluster members and the field counterparts are shown in the middle and right panel of Figure \ref{f_vpd_bin}, respectively. Clearly, the PM selection is still not perfect because it leaves some blue foreground disk stars, but the narrow red giant branch (RGB) of the cluster is well defined. 

\begin{figure}[h]
\plotone{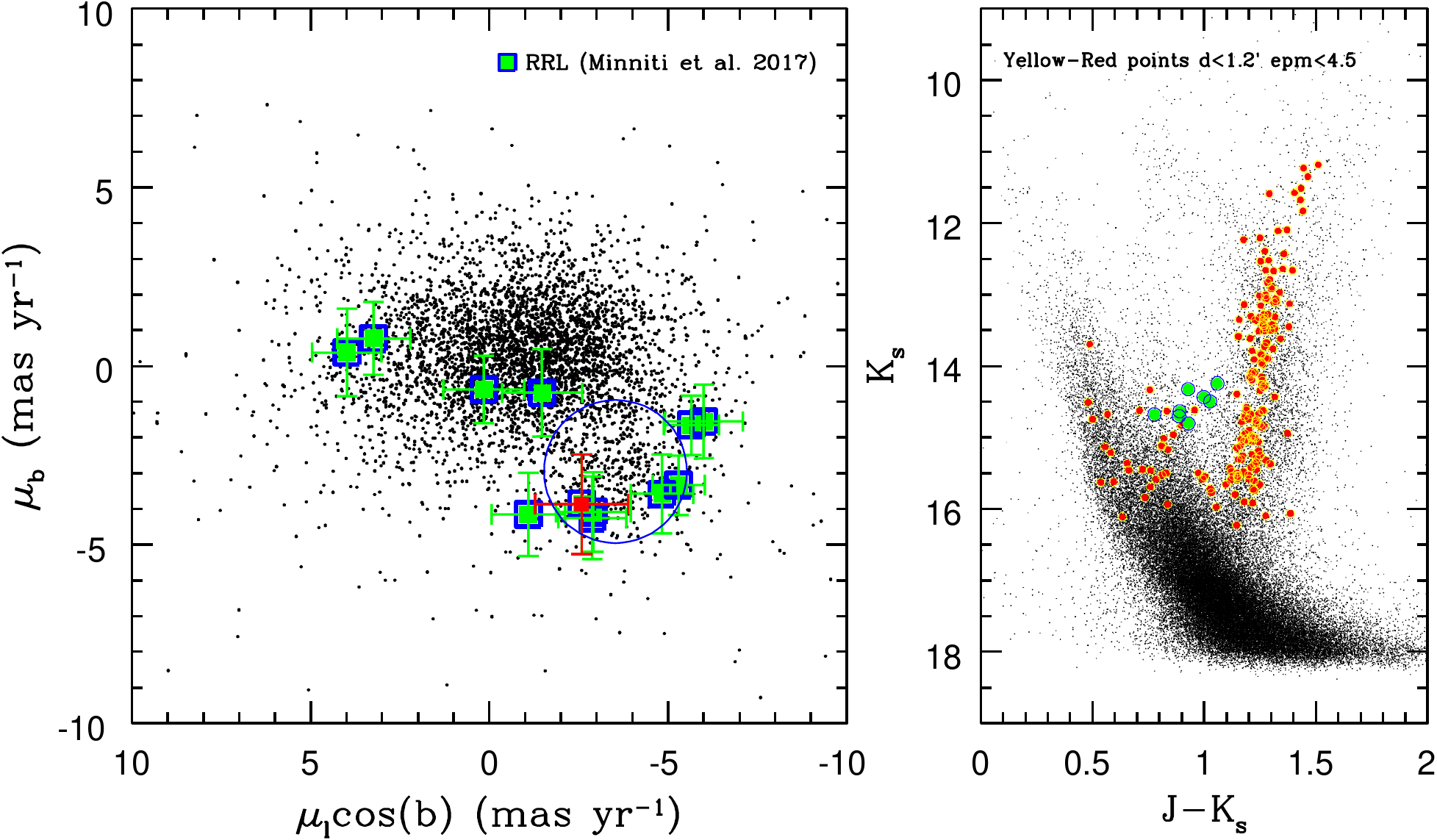}
\caption{Left panel: VPD  for all the stars with $K_s<15$ mag. The green-blue squares indicate the position of the 12 RRL discovered by \cite{minniti17_1}. The blue circle, with a radius of two mas~yr$^{-1}$, shows the adopted criterion used to separate cluster members from disk interlopers. As can be seen, nine RRL fall, within the errors, inside the adopted region and likely belong to VVV-GC05. The outermost RRL member (d025-0114911) is highlighted in red. Right panel: VVV near-IR CMD  for all the detected sources in our study. Stars falling inside the blue circle in the left panel with relatively low associated statistical errors, and closer than 1.2 arcmin from the cluster center are represented by yellow-red dots, while the confirmed RRL stars are shown with green-blue circles. A number of stars in the blue region suggest that the HB of VVV-GC05 extends blueward of the instability strip.}  
\label{f_combinada}
\end{figure}

\cite{minniti17_1} found a dozen RRL likely associated with VVV-GC05, eight of them being tightly packed in the inner region of the cluster. With the PMs in hand, we can now firmly establish their cluster membership. Our results are presented in Figure~\ref{f_combinada} (left panel), where we include, for the sake of clarity, only stars with $K_s<15$. We indicate in green-blue the RRL stars from \cite{minniti17_1}, and with the blue circle the area containing the more likely cluster members. 
From the analysis of Figure \ref{f_combinada}, three RRL are likely field interlopers (d025-0157039, d025-332556, and d025-0175388) and nine (five type RRab plus four type RRc) RRL can be considered, within the errors, bona-fide cluster members. We performed a new distance measurment for the cluster using this new RRL sample. We adopted a mean intrinsic color of ($J-K_{s})_0=0.17\pm0.03$ for the RRL type ab \citep{navarrete15} to derive a cluster reddening value of $E(J-K_s)=0.8$ mag, which corresponds to $A_{k_s}=0.42$ using the extintion ratio $R=0.528$ of \cite{nishiyama09}. Applying equation (17) from \cite{muraveva15} and adopting a metallicity ${\rm [Fe/H]}=-1.4$ dex \citep{minniti17_1,koch17}, the mean distance modulus of the nine RRL members is $(m-M)_0=14.38$, equivalent to 7.5 kpc, in perfect agreement with the value obtained by \cite{minniti17_1}. Interestingly, among the confirmed RRL members, one of them (d025-0114911) is located at $\sim 11$ arcmin from the cluster center, well beyond the measured cluster extension \citep[$\sim 3$ arcmin,][]{minniti17_1}. This corresponds to a projected separation of 6.5 pc at the distance of $D=7.5$ kpc, suggesting that VVV-GC05 may be losing stars due to the gravitational interaction with the Milky Way (MW) and/or cluster evaporation. Based on our new sample of RRL, we also recomputed the mean period of its fundamental-mode stars, obtaining  $\langle P_{ab} \rangle$=0.574 days. This result confirms the classification as an Oosterhoff I GC (\citeauthor{oosterhoff39} \citeyear{oosterhoff39}; 
see \citeauthor{catelansmith15} \citeyear{catelansmith15}, for a recent review and references).


In order to obtain a representative and clean sample of objects likely belonging to VVV-GC05, we selected stars with very similar kinematics (inside the blue circle shown in Figure~\ref{f_combinada}) with relatively small PMs statistical errors ($\rm{epm}=\sqrt{\rm{se}^2_{\mu_{l}\cos(b)}+\rm{se}^2_{\mu_{b}}}<4.5$ mas~yr$^{-1}$), and with similar projected position in the sky (within 1.2 arcmin from the cluster center). The selected stars are shown in the right panel of Figure~\ref{f_combinada}, overplotted in the CMD containing all the stars detected in the studied field. The inspection of the CMD shows a well defined RGB and a clear overdensity of stars at $J-K_s=1.30 \pm 0.05$, $K_s=13.35 \pm 0.05$, that we identified as the RGB bump. We include in green-blue the PM confirmed RRL members. The variability campaign in the VVV survey is only conducted with the $K_s$ filter, and therefore colors arise from only a few $J$ images (two in this case). As expected, because of the decoupling of their $K_s,J$ magnitudes some of these variables may fall outside the instability strip. The presence of cluster members in the blue plume region suggests that the HB of VVV-GC05 may extend significantly to the blue through a blue tail that reaches $J-K_s=0.5,K_s=16$. However we cannot rule out that contamination from the disk is still present. The mere presence of blue HB stars (in addition to RR Lyrae variable stars) also confirms that the cluster is older than 10 Gyr \citep[see, e.g.,][]{catelan09}.

Finally, based on the most likely star members, we attempt to measure the shape of the cluster. As shown in Figure \ref{f_flat}, the cluster core is fairly spherical, with the flattening being considerable only in the outskirts, along the Galactic longitude. From the outermost contours we can visualy measure a shape of $3.2^{\prime} \times 4.5^{\prime}$, giving  an overall oblateness $b/a \approx 0.7$, or in the notation of \cite{hubble36} $e=(a-b)/a \approx 0.18$, ranking VVV-GC05 as one of the most flattened GCs known \citep[comparing with the lists of][]{white87,chen10}. \cite{vandenbergh08} argued that many of the most flattened GCs suffer from differential reddening (citing the most flattened cluster M19 = NGC~6273 as the prime example), however, since our measurements \citep[like those of][]{chen10} are in the near-IR, this problem is minimized. Moreover, the dispersion of the VVV-GC05 RGB colors discard that differential reddening is a significant effect, and cannot explain the observed flatness. We take this significant elongation as evidence that the cluster is being dynamically distorted by the Galactic field. However, the caveat is that these shape measurements are difficult to interpret not only because of the dense stellar background and differential reddening, but also because of the presence of a couple of bright stars in the immediate vicinity of the cluster. Also note that there are other clusters with radial variations in the ellipticity profile as well, most notably $\omega$ Cen \citep{geyer83,anderson10}. This suggests that tidal effects are not the only mechanism that can produce such variations. 

\section{THE VVV-GC05 RADIAL VELOCITIES} 
\label{rv}
In order to obtain radial velocities for cluster red giant stars, we used the near-IR FIRE spectrograph (${\rm R \sim 6000}$) at the Magellan Baade 6.5m Telescope at Las Campanas Observatory. 
The date of the spectroscopic observations was June 3rd, 2017, with excellent seeing conditions (sometimes below 0.4 arcseconds). We managed to obtain high quality FIRE spectra for 6 cluster candidate red giant stars that were proper motion selected. The observations of these 6 red giant stars took about 1.3 hours, which is relatively fast because they are bright and close to each other.
The FIRE spectra show spectral order with strong curvature, seriously complicating the reduction process.
The reduction of these spectra was done with our own Python reduction cascade, that 
includes bias subtraction, stray-light modeling, flat fielding, correction for inhomogeneous slit illumination, cosmic ray/bad pixel rejection, sky subtraction, optimal extraction, telluric correction, and wavelength calibration. In doing so we have cut the three orders in the bluest side of the spectra because of problems with the order rectification.
For the blaze function and telluric correction, we obtained a spectrum of the telluric standard HD141542 (A0V), and followed the correction method outlined in \cite{vacca03}. However, having just one telluric spectrum at the beginning of the observations, we cannot map the temporal changes in the tellurics and therefore the fluxes in the corrected observed spectrum at the location of those telluric lines are accurate only within a few 10\% (which should not affect the radial velocities).
For the wavelength calibration we used sky emission lines in the science exposures to fit a global wavelength solution to all the orders, and then used our own  cross-correlation code to measure radial velocities. After Fourier filtering as well as tellurics --and sky residuals-- masking of the pseudo normalized spectra, the code computes a weighted normalized cross correlation function for all the spectra in the APOGEE H-band template library, convolved with a Gaussian in order to appropriately match the resolution of FIRE ($\sim50$ km/s). We use the mean barycentric velocity of the peaks of two FIRE orders corresponding to FIRE's H-Band coverage. Subsequently, we fitted the individual cross-correlated profiles and deduced RVs from the positions of the highest peaks corresponding to the templates with the best matches to the observations. We employed the same method to spectra of Gaia radial velocity standards \citep{soubiran13} taken during the same observing run, and could not find significant deviations from the literature values. Hence the applied method should be reliable.

\begin{figure}[h]
\plotone{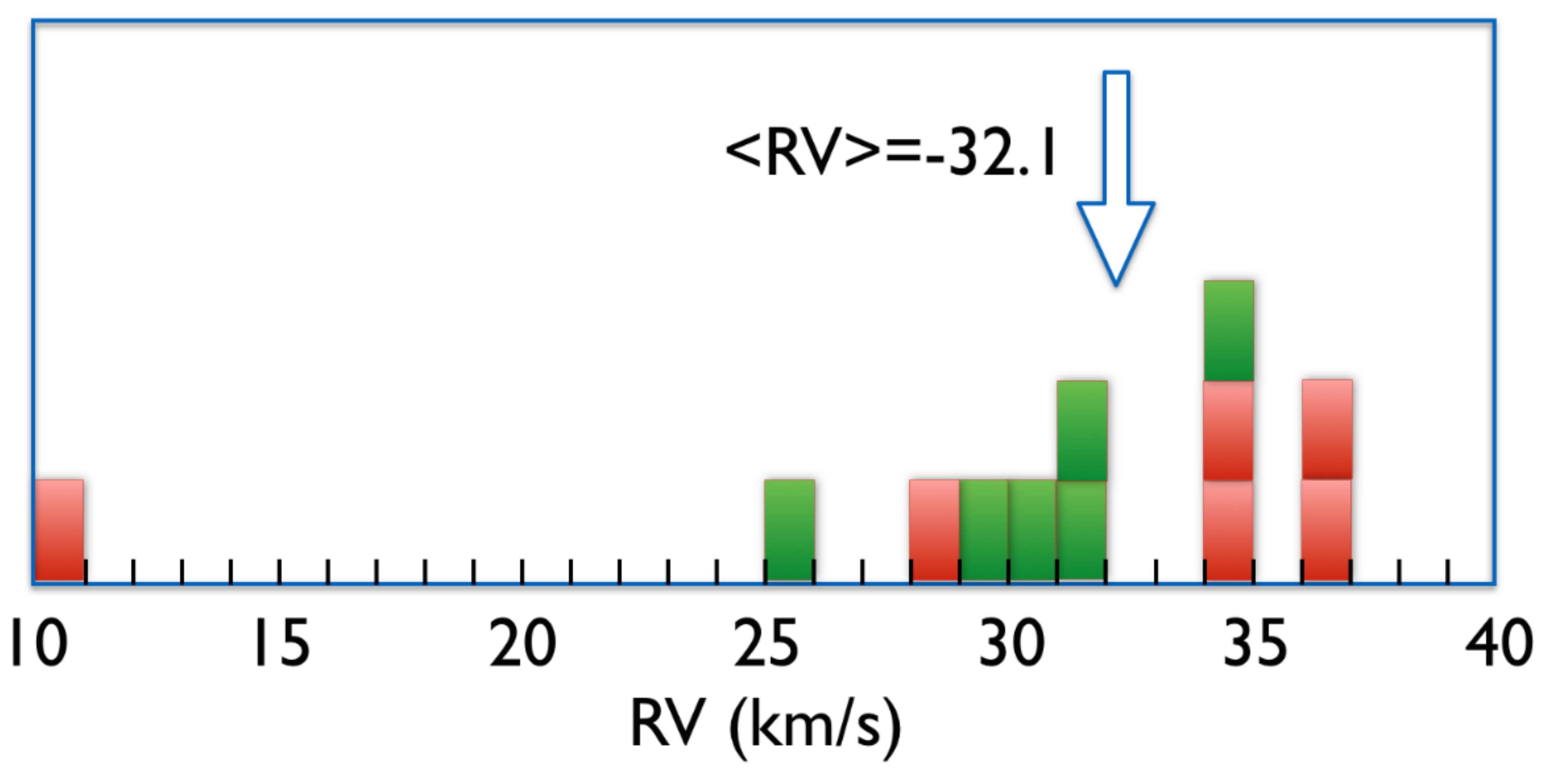}
\caption{Radial velocities for red giants in the field of the globular cluster VVV-GC05, showing the velocities measured by \cite{koch17} in green, and those measured in this work in red.} 
\label{f_rv}
\end{figure}

After measuring the RVs of the six proper motion selected cluster candidates (listed in Table \ref{rad_vel}), we conclude that 5 stars are cluster members 
and one star is a clear outlier (star d025-305014339, which is also the brightest sample star). Comparing the respective spectra, we see that this outlier star seems to be a much more metal-rich foreground star, with strong lines, while the other five are fairly similar and metal-poor, with a well defined continuum and weak lines. \cite{koch17} also measured radial velocities for six red giant members of this cluster. Their radial velocities are in reasonable agreement with ours, as shown in Figure \ref{f_rv}, although there is a small offset between their mean velocities and ours, $\Delta RV = 3.7$ km/s, that is not statistically significant ($2\sigma$). We measure a mean cluster radial velocity of RV$=-34.0 \pm 1.5$ km/s based on five red giant members, and \cite{koch17} measured a mean cluster radial velocity of RV$=-30.3 \pm 1.2$ km/s based on six red giant members.  We measure a velocity dispersion of $\sigma=2.9 \pm 0.9$ km/s, slightly larger than that measured by \cite{koch17}, $\sigma=2.5 \pm 0.9$ km/s. Also, because of the small difference in the mean velocities of the two samples, the combined velocity dispersion is somewhat larger: $\sigma=3.4 \pm 0.7$ km/s. This larger velocity dispersion value would almost double the mass of this cluster with respect to that estimated by \cite{koch17}, who found a total dynamical mass of $M = 1.4 \times 10^4$ $M_{\odot}$. Clearly, more radial velocities need to be measured in order to definitely establish the cluster radial velocity dispersion and total mass. In order to compute the cluster orbital properties, hereafter we will adopt the unweighted mean of these two independent determinations for the cluster radial velocity,
RV$=-32.1 \pm 1.5$ km/s. 

\section{CHARACTERIZATION OF THE GLOBULAR CLUSTER ORBIT} 
\label{orbit}
In order to compute the GC orbit, we need absolute PMs instead of relative ones. The transformation from the relative to the absolute system was done using more than a thousand stars in common between our reference stars for computing PMs (see section \ref{pm}) and the Gaia DR2 catalog recently published by \cite{gaia18}. We cross-correlated both catalogs allowing a maximum difference of one arcsec projected distance. From their comparison, we computed a mean relative shift of $-4.95\pm0.04$ mas~yr$^{-1}$ along Galactic longitude and $-0.76\pm0.04$ mas~yr$^{-1}$ along Galactic latitude. The derived absolute PM for VVV-GC05 resulted in ($\mu_{l}\cos(b),\mu_{b}$)=($-8.69\pm0.1,-3.62\pm0.1$) mas~yr$^{-1}$. To trace the orbit, we use the parameters listed in Table \ref{gc_data}. For reference, the 3-dimensional solar velocity and velocity of the local standard of rest adopted in this work is: [U$_{\odot}$,V$_{\odot}$,W$_{\odot}$] = [11.1, 12.24 , 7.25] km/s  \citep{schonrich10}, the Sun is located at R$_{\odot}$ = 8.3 kpc, and the local rotation velocity is assumed to be $\Omega_{0}$(R$_{\odot}$) = 239 km/s, given by \cite{sofue15}. All these input quantities are well determined, and their errors are adequate for obtaining good orbital parameters. 

We performed a numerical integration of orbits for FSR1716 in a barred Milky Way model, adopting a new modern galaxy algorithm called GravPot16 \citep{ftrincado17_2}, that makes an important effort to fit specially the inner Galactic region. The parameters employed for the Galactic potential are listed in Table \ref{parameters}. GravPot16 is a semi-analytical, steady-state, 3D gravitational potential of the Milky Way, observationally and dynamically constrained. The model is primarily made up of the superposition of several composite stellar components, where the density profiles in cylindrical coordinates, $\rho_{i}$(R, z), are the same as those proposed in \cite{robin03,robin12,robin14}, i.e., a boxy/peanut bulge, a Hernquist stellar halo, seven stellar Einasto thin disks with spherical symmetry in the inner regions, two stellar sech$^{2}$ thick disks, a gaseous exponential disk, and a spherical structure associated with the dark matter halo. A new formulation for the global potential, $\Phi$(R, z), of this Milky Way density model will be described in detail in a forthcoming paper (Fernandez-Trincado et al. 2018, in preparation). 
We consider an angle of $\phi$ = 20$^{\circ}$ for the present-day orientation of the major axis of the Galactic bar and the Sun-Galactic center line. The total mass of the bar taken in this work is 1.1$\times$10$^{10}$ M$_{\odot}$, that corresponds to the dynamical constraints towards the Milky Way bulge from massless particle simulations \citep[e.g.,][and Fern\'andez-Trincado et al. 2018, in preparation]{ftrincado17_2}, which is consistent with the recent estimate given by \cite{portail17}. Moreover the bar potential model has been computed adopting a new mathematical technique which considers ellipsoidal shells with similar densities. Additionally, it should be noted that GravPot16 has been extensively employed to predict stellar orbits and/or orbital parameters for a large set of sources \citep[see][]{ftrincado16,ftrincado17_3,ftrincado17_2,ftrincado17_1,anders17,abolfathi18}. 

\begin{figure}[h]
\plotone{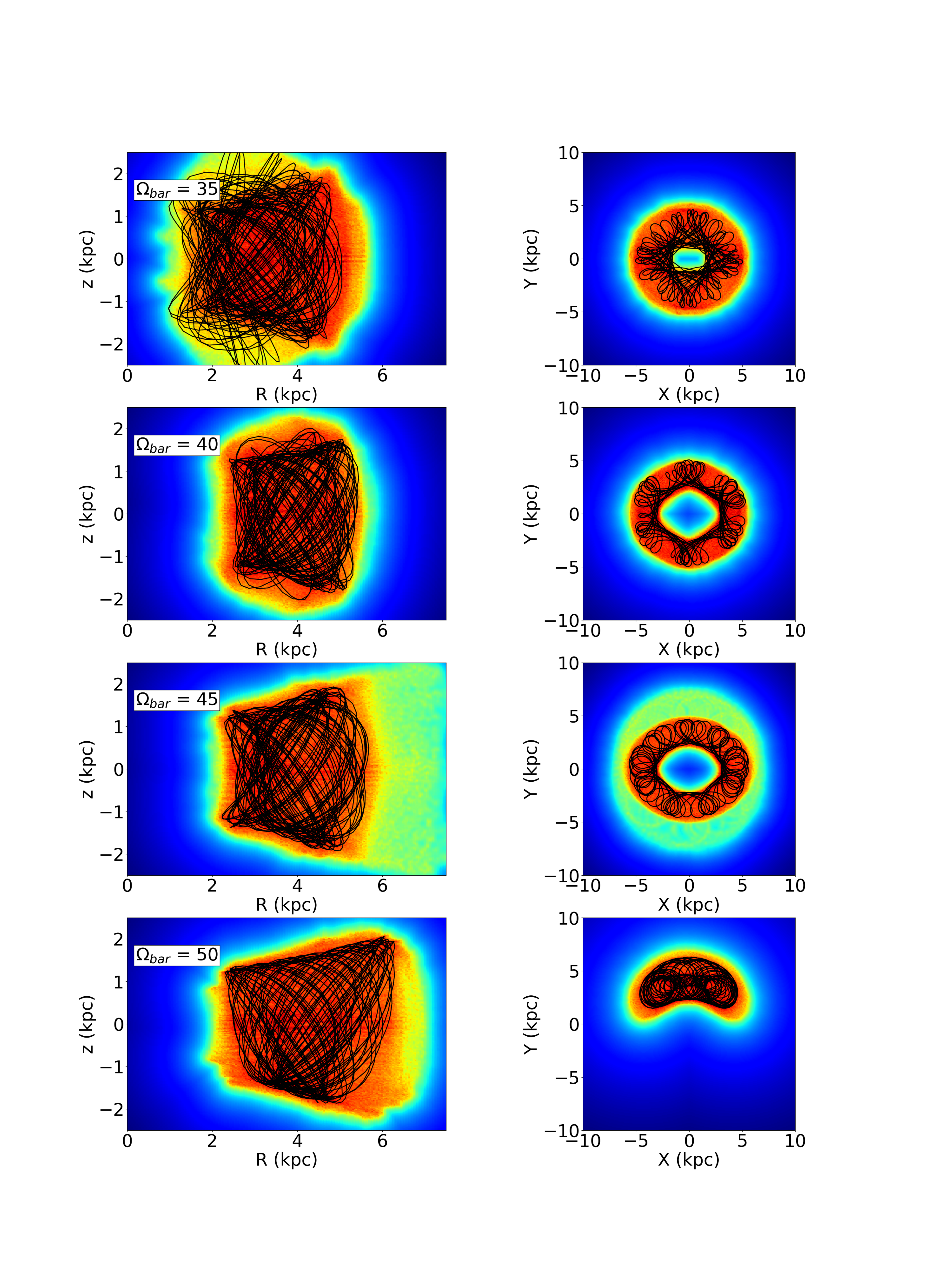}
\caption{Probability density in the meridional Galactic plane (left column) and equatorial Galactic plane (right column) of one million simulated orbits of FSR1716 time-integrated backwards for 2.5 Gyr in the non-inertial reference frame where the bar is at rest. The internal label indicate the adopted pattern speed of the bar. The black path shows the nominal orbit while red and yellow colors correspond to more probable regions of the space, which are crossed more frequently by the simulated orbits.}
\label{f_orbit}
\end{figure}

The orbit of FSR1716 is computed adopting a simple Monte Carlo procedure for different bar pattern speeds as mentioned above. For each simulation, we time-integrated backwards one million orbits for 2.5 Gyr under variations of the initial conditions (proper motions, radial velocity, heliocentric distance, Solar position, Solar motion and the velocity of the local standard of rest) according to their estimated errors, where the errors are assumed to follow a Gaussian distribution. 
It is important to note that the uncertainties in the orbital predictions are primarily driven by the uncertainty in the PMs and distances with a negligible contribution from the uncertainty in the radial velocity. In Figure~\ref{f_orbit} we show the meridional and equatorial orbit of FSR1716 with the four different values of the angular velocity of the Galactic bar (different rows). Small differences can be underlined between the different orbits. The radial excursions of the orbit of FSR1716 seem to increase slightly with the bar pattern speed, while the vertical excursions remain similar in these cases. The less frequent closer approaches to the innermost part of the Galaxy imply highly likely a larger survival expectancy for the cluster. 

We see that the ${\rm R_{min}}$ appears to increase and decrease periodically depending on the bar angular velocity, and speculate that this means that the tidal destruction rate of Galactic GCs in the innermost parts of the Milky Way would increase and decrease periodically as well. This calls for further dynamical modelling for this cluster and for the search of similar clusters that would test the tidal effects in the central parts of our galaxy.

The main results are listed in Table \ref{results}. The orbit results to be confined within ${\rm |Z_{max}|<2.0}$ kpc, and has a significant eccentricity $0.4<e<0.6$, with perigalactic distance ${\rm 1.5<R_{peri}(kpc)<2.3}$, and apogalactic distance ${\rm 5.3<R_{apo}(kpc)<6.4}$. These orbital parameters are consistent with a GC belonging to the inner halo \citep[e.g.][]{moreno14, bobylev17}. Some limitations of our calculations include not considering the effects of dynamical evolution of the cluster and tidal effects on it exerted by the Galactic disk in each orbit, mass loss, and secular changes in the Milky Way potential over time.

\section{DISCUSSION: DYNAMICAL DISRUPTION PROCESSES IN ACTION?} 
\label{discussion}

Significant flattening (elliptic shapes) have been observed in some GCs \citep{white87}, although not necessarily disk shocking is the main process acting to flatten most GCs \citep{shimada98}. The presence of rotation can significantly flatten a GC, or in some cases differential reddening can account for the elliptical shapes \citep{vandenbergh08}, as confirmed by \cite{chen10}, who found rounder values based on near-IR photometry from 2MASS. However, \cite{stephens06} also found at least one case in which high GC flattening may be caused by velocity anisotropy. 

Disk shocking is due to the varying tidal field as a GC plunges through the MW disk, and has been proposed as an important destruction mechanism depending on the cluster location and orbit \citep{weinberg94}. According to early theoretical expectations, gravitational disk shocking would cause the expansion and subsequent mass loss for such a cluster \citep{weinberg94}. This effect would be more dramatic for clusters with smaller Galactocentric distances than ${\rm R_{GC} \sim 7.5}$ kpc, which is the case of VVV-GC05 that is located at ${\rm R_{GC}=4.3}$ kpc \citep{minniti17_1}.  To test the disk shocking effect it was needed a GC where all the other dynamical effects were reduced. So the cluster has to be located in the MW plane but away from the bulge to avoid bulge shocking, and be massive enough to withstand evaporation. VVV-GC05 is apparently such a cluster, located at $b=-1.5$ deg, and height $Z = 209$ pc below the Galactic plane. In this work we have determined the orbit of this GC, finding that is presently plunging through the Galactic disk, which increases the effect of disk shocking. Therefore, this GC is a laboratory like no other to test the effects of disk shocking. 



\begin{figure}[h]
\plotone{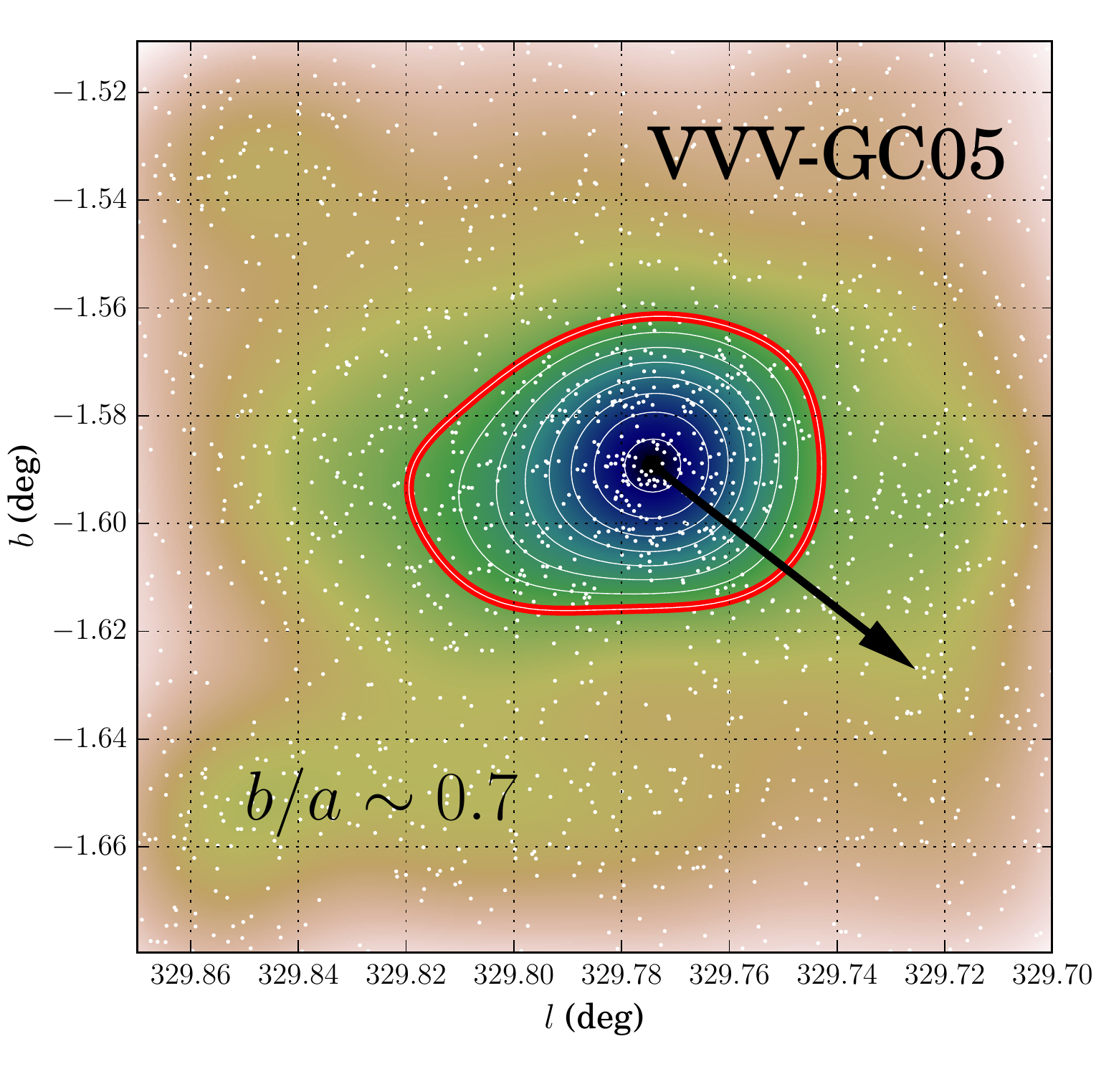}
\caption{Surface density distribution of PM selected stars in the field of the cluster. The central region is spherical, while the outer parts are significantly flattened along the Galactic longitude coordinate. The tidal extensions to both sides of the cluster are evident in this graph. The quoted value of the cluster's elongation was visually estimated from the density contour highlight. The black arrow points along the direction of the cluster orbit. For visual aid, spatial density is color-coded in arbitrary linear units.} 
\label{f_flat}
\end{figure}

In principle, this cluster is subject to severe disk shocking, which can be the cause for the elongated shape observed in its outer parts (Figure \ref{f_flat}). The evaporation due to tidal heating should expel stars that could be found a few tidal radii away from the clusters \citep[e.g.][]{meylan02}. Interestingly, one RRL found by \cite{minniti17_1} is located at a distance $>11$ arcmin from the cluster centre, well beyond its spatial extension (radius of $\sim 3$ arcmin), suggesting that this may be the case, but the numbers are low. 
Another piece of evidence pointing at the existence of tidal tails is the difficulty found when trying to decontaminate the cluster CMD using statistical background decontamination methods \citep{minniti17_1}. The PM selected cluster members show an extended distribution elongated along the Galactic longitude coordinate (Figure \ref{f_flat}), which resembles tidal extensions on both sides of the cluster.

Unfortunately, one major uncertainty that remains to be settled is the total mass of the cluster. As discussed in the previous section \ref{rv}, \cite{koch17} measured a total dynamical mass of $M= 1.4 \times 10^4$ M$_\odot$ based on a velocity dispersion of $\sigma=2.5 \pm 0.9$ km/s from six stars, while the velocity dispersion measured here using the total combined sample of 11 stars deemed to be cluster members would ($\sigma=4.5 \pm 1.5$ km/s) almost doubles the mass of this cluster. 

Orbit and mass would provide important constrains to address several interesting questions about VVV-GC05: Did this cluster belong to an accreted satellite? Does its present location reflect the result of spiraling into the Milky Way by dynamical friction? Is this cluster presently in the final stages of disruption? Has this cluster lost most of its mass? Unfortunately, its location at low Galactic latitude where the stellar density is maximum prevents us from answering these questions yet. 

\section{CONCLUSIONS}

We have measured the mean radial velocities and PMs for VVV-GC05, obtaining an orbit for the cluster. The computed orbit discards the hypothesis that this is a disk GC, in spite of its special location in the Galactic plane. We conclude that this is a metal-poor GC of the inner halo, that has survived in an elliptical orbit confined to ${\rm 1.5<R_{GC}(kpc)<6.4}$. 

We argue that this cluster is presently plunging through the MW plane and therefore is being severely affected by disk shocking.
The tentative detection of tidal extensions on the spatial distribution of RR Lyrae and of PM selected member stars of this GC support our conclusions.

This target is strategically located at $b=-1.5$ deg in the MW disk (with $Z = 209$ pc below the plane), and it is therefore specially important to test  the effects of dynamical evolution of Galactic GCs. Indeed, detailed models in combination with careful observations of VVV-GC05 may be able to directly measure the force field in that special place of the MW at ${\rm R_{GC}=4.3}$ kpc.

\acknowledgments
We gratefully acknowledge data from the ESO Public Survey program ID 179.B-2002 taken with the VISTA telescope, and products from the Cambridge Astronomical Survey Unit (CASU). Support is provided by the BASAL Center for Astrophysics and Associated Technologies (CATA) through grant PFB-06, and the Ministry for the Economy, Development and Tourism's Iniciativa Cient\'ifica Milenio grant IC120009, awarded to the Millennium Institute of Astrophysics (MAS). D.M., J.G.F-T, M.Z., M.C., and J.A.G. acknowledge support from FONDECYT grants Regular No. 1170121, 3180210, 1150345, 1171273, and Iniciacion 11150916, respectively. F.G. acknowledge support from CONICYT-PCHA Doctorado Nacional 2017-21171485 and Proyecto Fondecyt Regular 1150345. A.P-V acknowledges FAPESP for the postdoctoral fellowship 2017/15893-1. GH acknowledges support from the Graduate Student Exchange Fellowship Program between the Institute of Astrophysics of the Pontif\'icia Universidad Cat\'olica de Chile and the Zentrum f\"ur Astronomie of the University of Heidelberg, funded by the Heidelberg Center in Santiago de Chile and the Deutscher Akademischer Austauschdienst, and by CONICYT-PCHA/Doctorado Nacional grant 2014-63140099. We are grateful to the Aspen Center for Physics where our work was supported by National Science Foundation grant PHY-1066293, and by a grant from the Simons Foundation (D.M. and M.Z.). J.G.F-T gratefully acknowledges funding for the GravPot16 software provided by the Centre national d'\'etudes spatiale (CNES) through grant 0101973 and UTINAM Institute of the Universit\'e de Franche-Comte, supported by the Region de Franche-Comte and Institut des Sciences de l'Univers (INSU).
\\

\bibliography{biblio}


\begin{figure}[h]
\plotone{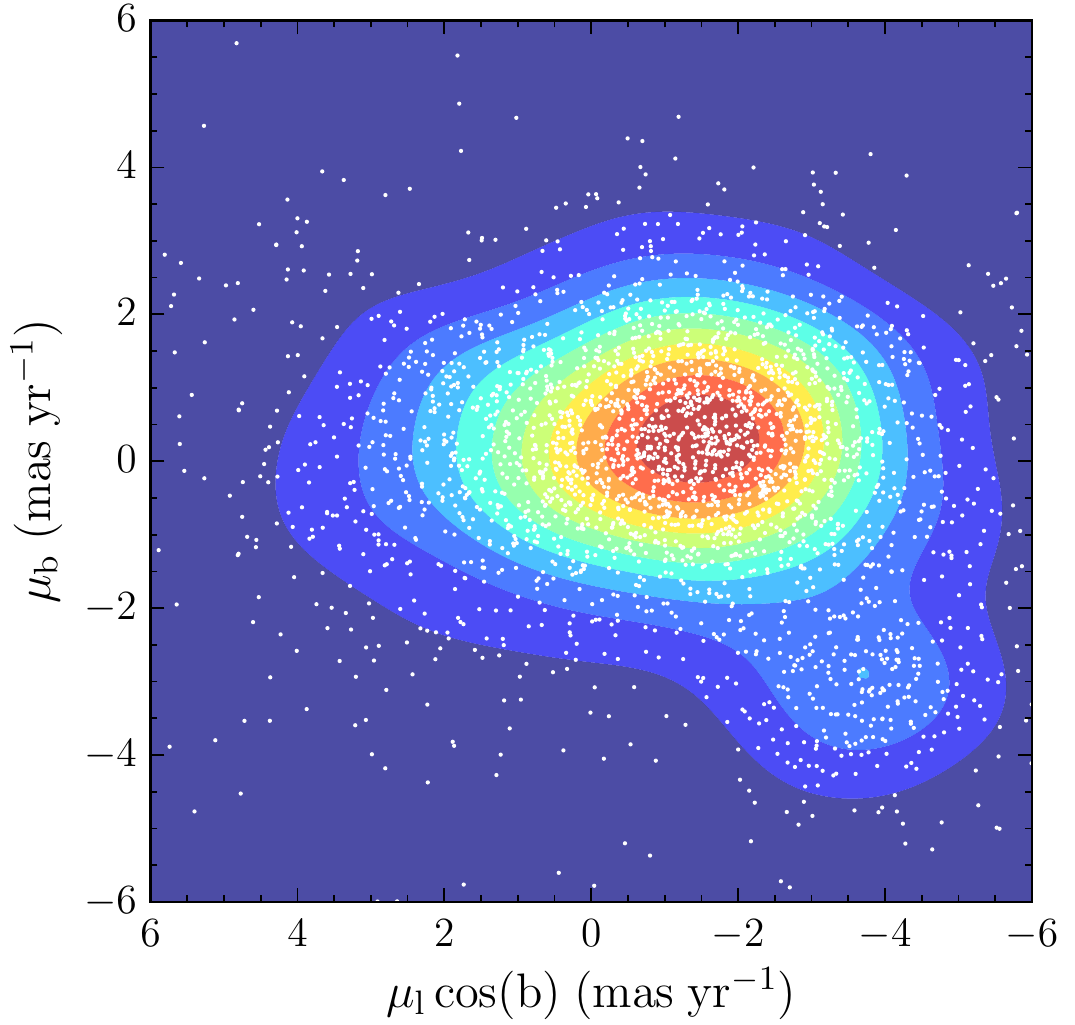}
\caption{Smooth VPD for all the stars with $K_s<15$ mag. Likely stars belonging to VVV-GC05 in the lower right corner clearly separate from the disk stars. The cloud of disk stars in the center appears elongated due to the disk rotation. For visual aid, spatial density is color-coded in arbitrary linear units.} 
\label{f_vpd}
\end{figure}

\begin{deluxetable}{ccccccccc}
\tablecaption{Radial Velocity Measurements.\label{rad_vel}}
\tablehead{
\colhead{ID} & \colhead{RA2000} & \colhead{DEC2000} & \colhead{RV (km/s)} & \colhead{$\sigma_{RV}$ (km/s)} & \colhead{$K_s^a$}  & \colhead{$J^a$} & \colhead{$J-K_s$}\\
}
\startdata
d025-305014339& 242.63448& -53.71523& -10.39 &  0.62&  9.542& 10.935& 1.393\\
d025-304820647& 242.76954& -53.75650& -36.85 &  1.89&  9.845& 11.333& 1.488\\
d025-305435610& 242.60974& -53.76977& -28.62 &  2.12& 10.020& 11.394& 1.374\\
d025-304822561& 242.75303& -53.70405& -36.22 &  1.43& 10.060& 11.561& 1.501\\
d025-305013143& 242.58673& -53.72942& -34.04 &  2.22& 11.184& 12.599& 1.415\\
d025-304816009& 242.64867& -53.78362& -34.40 &  2.06& 11.402& 12.841& 1.439\\
\enddata
\tablenotetext{a}{Typical photometric errors are $\sigma_{K_s} = 0.03$ mag, and $\sigma_{J} = 0.01$ mag. }
\end{deluxetable}

\begin{deluxetable}{lc}
\tablecaption{Observational Data for VVV-GC05.\label{gc_data}}
\tablehead{
\colhead{ } & \colhead{Value} \\
}
\startdata
$(\alpha, \delta)_{J2000}$&$(16:10:30.0,-53:44:56) $  \\
$(l,b)$&$(329.77812^\circ,-1.59227^\circ) $ \\
$d_\odot$& $7.5\pm0.2$ kpc \\
$v_r$& $-32.1\pm 1.5$ km/s\\
$\mu_l \cos b$& $-8.69\pm 0.1$ mas yr$^{-1}$\\
$\mu_b$& $-3.62\pm 0.1$ mas yr$^{-1}$\\
\enddata
\end{deluxetable}

\begin{deluxetable}{lcr}
	\tablecolumns{3}
	\tablewidth{0pt}
	\tablecaption{Parameters employed for the Galactic potential.\label{parameters}}
	\tablehead{\colhead{Parameter} &\colhead{Value} & \colhead{References}}
	\startdata
	${\rm R_0}$               & 8.3$\pm$0.23 kpc     & 1 \\
	${\Omega}_0$        & 239$\pm$7 km/s       & 1 \\
	${\rm(U,V,W)_{\odot}}$   & (-11.1$^{+0.69}_{-0.75}$, 12.24$^{+0.47}_{-0.47}$, 7.25$^{+0.37}_{-0.36}$) & 2 \\
	\cutinhead{Galactic Bar}
	Present position of major axis              & 20$^{\circ}$        & 3,4 \\
	Angular velocity                    & 35,40,45,50 km/s kpc$^{-1}$ & 3,4,5,6,7 \\
	Cut-off radius     &  3.28 kpc & 8\\                   
	\enddata
	\tablerefs{
		1)~\citet{brunthaler11}.
		2)~\citet{schonrich10}.
		3)~\citet{ftrincado17_1}.
		4)~\citet{ftrincado17_2}.
		5)~\citet{portail15}.
		6)~\citet{monari17_1}.
		7)~\citet{monari17_2}.
		8)~\citet{robin12}.
	}
\end{deluxetable}


\clearpage
\begin{deluxetable}{cccccc}
	\tablecaption{Monte Carlo orbital parameters of FSR1716. The average value of the orbital parameters was found for  the one million realizations, with uncertainty ranges given by the 16th (subscript) and 84th (superscript) percentile values.\label{results}}
	\tablewidth{0pt}
	\tablehead{
		\colhead{${\Omega}_{\rm bar}$} &  \colhead{$\langle$$r_{\rm peri}$$\rangle$} &
		\colhead{$\langle$$r_{\rm apo}$$\rangle$} & \colhead{$\langle$$|z|_{\rm max}$$\rangle$} & 
		\colhead{$\langle$$e$$\rangle$}  \\ 
		\colhead{(km/s kpc$^{-1}$)} & \colhead{(kpc)} & \colhead{(kpc)} & \colhead{(kpc)} & \colhead{}
	}
	\startdata
        35 & 1.50$^{2.49}_{1.09}$ &  5.29$^{5.56}_{5.09}$ & 2.09$^{2.64}_{1.82}$  &  0.57$^{0.65}_{0.35}$  \\ 
        40 & 2.31$^{2.41}_{2.22}$ &  5.42$^{5.51}_{5.32}$ & 2.11$^{2.23}_{1.99}$  &  0.40$^{0.42}_{0.38}$  \\ 
        45 & 2.24$^{2.34}_{2.14}$ &  5.75$^{5.91}_{5.53}$ & 2.00$^{2.07}_{1.91}$  &  0.44$^{0.46}_{0.42}$  \\ 
        50 & 2.29$^{2.43}_{2.09}$ &  6.39$^{6.75}_{6.19}$ & 2.02$^{2.12}_{1.89}$  &  0.47$^{0.53}_{0.44}$  \\ 
	\enddata
\end{deluxetable}


\end{document}